\documentclass[12pt,english]{article}

\usepackage{geometry}
\usepackage{color}
\usepackage{babel}
\usepackage{amsmath}
\usepackage{amsthm}
\usepackage{amssymb}
\usepackage[linesnumbered,ruled,vlined]{algorithm2e}
\usepackage{graphicx}
\usepackage{setspace}
\usepackage[longnamesfirst,sort,comma]{natbib}
\usepackage[unicode=true,pdfusetitle,
 bookmarks=true,bookmarksnumbered=false,bookmarksopen=false,
 breaklinks=false,pdfborder={0 0 1},colorlinks=true]{hyperref}
\hypersetup{linkcolor=red, citecolor=blue}
\usepackage{enumitem}
\usepackage[all]{hypcap}
\usepackage{breakcites}

\hypersetup{
           breaklinks=true,   
           colorlinks=true,   
           pdfusetitle=true,  
        }

\theoremstyle{plain}
\newtheorem{thm}{\protect\theoremname}[section]
\theoremstyle{plain}

\ifx\proof\undefined
\newenvironment{proof}[1][\protect\proofname]{\par
	\normalfont\topsep6\p@\@plus6\p@\relax
	\trivlist
	\itemindent\parindent
	\item[\hskip\labelsep\scshape #1]\ignorespaces
}{%
	\endtrivlist\@endpefalse
}
\providecommand{\proofname}{Proof}
\fi
\theoremstyle{remark}

\theoremstyle{definition}

\theoremstyle{plain}

\newtheorem{hp}{Assumption}

\newtheorem{definition}{Definition}
\newtheorem{cor}{Corollary}[section]

\providecommand{\definitionname}{Definition}
\providecommand{\lemmaname}{Lemma}
\providecommand{\propositionname}{Proposition}
\providecommand{\remarkname}{Remark}
\providecommand{\theoremname}{Theorem}

\def\1{\mathbb{I}}

\setcitestyle{citesep={;}}

\renewcommand{\baselinestretch}{1.2}

\SetCommentSty{mycommfont}

\SetKwInput{KwInput}{Input}                
\SetKwInput{KwOutput}{Output}              

\begin{document}
\title{Partly Linear Instrumental Variables Regressions without Smoothing on the Instruments \thanks{We would like to thank Pascal Lavergne for inspiring discussions. We are also grateful to Juan Carlos Escanciano, Ingrid Van Keilegom, Jad Beyhum, Valentin Patilea, and Xavier d'Haultfoeuille for their comments and suggestions. Jean-Pierre Florens acknowledges funding from the French National Research Agency (ANR) under the Investments for the Future program (Investissements d'Avenir, grant ANR-17-EURE-0010).}
}

\author{Jean-Pierre  Florens \thanks{\emph{Toulouse School of Economics.
 Email: jean-pierre.florens@tse-fr.eu} Address correspondence: Toulouse School
 of Economics, 1 Esplanade de l'Universit\'e, 31080 Toulouse Cedex 06,
 FRANCE.} \hspace{.1cm} and 
\ Elia Lapenta\thanks{\emph{CREST and ENSAE. Email: elia.lapenta@ensae.fr} Address correspondence: CREST, 5 Avenue Le Chatelier, 91120 Palaiseau, FRANCE.} }

\date{September 15, 2023}

\maketitle
\thispagestyle{empty}
\vspace*{-.5cm}
\normalsize

\begin{abstract}
We consider a semiparametric partly linear model identified by instrumental variables. We propose an estimation method that does not smooth on the instruments and we extend the Landweber-Fridman regularization scheme to the estimation of this semiparametric model. We then show the asymptotic normality of the parametric estimator and obtain the convergence rate for the nonparametric estimator. Our estimator that does not smooth on the instruments coincides with a typical estimator that does smooth on the instruments but keeps the respective bandwidth fixed as the sample size increases. We propose a data driven method for the selection of the regularization parameter, and in a simulation study we show the attractive performance of our estimators.
\end{abstract}

\textbf{Keywords}: Instrumental Variables Regression, Partly Linear Model,  Ill Posed Inverse Problem, Landweber-Fridman Regularization.
\vspace{0.25cm}\\
\textbf{JEL Classification}: C01, C12, C14
\vspace{0.25cm}\\
\textbf{MSC Classification}: 45P05, 62G20, 62G08, 62G10, 62P20

\newpage
\setcounter{page}{2}

\section{Introduction}
Regressions with instrumental variables (IVs) play a central role in econometrics and have become increasingly popular in quasi-experimental studies. They are employed to recover causal effects and to estimate structural models suggested by economic theories. In this paper, we contribute to the literature on partly linear IV regressions by constructing an estimation method that does not smooth over the IVs and that relies on the Landweber-Fridman regularization. \\
We consider the partly linear model with endogenous regressors 
\begin{equation}\label{eq: PL NPIV model}
    Y=X^T\beta_0+ \phi_0(Z)+U\text{ with }\mathbb{E}\{U|W\}=0\, ,
\end{equation}
where $Y$ is a response variable, $X\in \mathbb{R}^\kappa$, $Z\in \mathbb{R}^p$,  $U$ is an unobserved error, and $W\in \mathbb{R}^q$ is a vector of instruments. Both $X$ and $Z$ are endogenous regressors, in the sense that the error $U$ is allowed to be correlated with them. The function $\phi_0$ is nonparametric. An empirical example giving rise to the above model is the estimation of the returns to schooling, where $Y$ represents the (log of the) wage of an individual, $X$ is her number of years of education, $Z$ is her work experience, and $U$ is an unobserved error containing the individual's unobserved ability. Both education and experience are endogenous, as they depend on the individual's unobserved ability. 
The IVs used to control for endogeneity are $W=(age,nearcollege)$, where $age$ represents the individual's age and $nearcollege$ is a proxy of the distance between the place where the individual grew up and an accredited four years college. See \cite{card1993using}. 

Our goals are (i) to propose an estimation method for the partly linear IV model in (\ref{eq: PL NPIV model}) that does not smooth over the IVs, (ii) to extend the Landweber-Fridman regularization to the estimation of the partly linear IV model, (iii) to obtain the convergence rate for the estimator of the nonparametric part of the model and the asymptotic normality for the estimator of the parametric part by the Landweber-Fridman regularization, and (iv) to draw a connection between our method and the classical method that smooths over the IVs. \\
Our first goal is thus to propose an estimation method for partly linear IV regressions that does not smooth over the IVs. Classical estimation methods such as \cite{darolles_nonparametric_2011} and \cite{florens_instrumental_2012} estimate $\beta_0$ and $\phi_0$ by applying the conditional expectation operator  $\mathbb{E}\{\cdot|W\}$ to both sides of (\ref{eq: PL NPIV model}). This gives rise to an integral equation, and the estimators of $\beta_0$ and $\phi_0$ are then built by taking the empirical counterpart of such an integral equation, see Section \ref{sec: A fixed bandwidth interpretation} for details. This, however, requires nonparametric estimation of the operator $\mathbb{E}\{\cdot|W\}$ and hence to smooth on the IVs. Differently, the method we propose in this paper avoids  smoothing on the instruments $W$. The main advantage of this is that we do not have to select a smoothing parameter for the IVs, see Section \ref{sec: Estimation by Landweber Fridman Regularization } for details.
\\
Our second contribution is to extend the 
Landweber-Fridman regularization scheme to the estimation of the partly linear IV model. The model in (\ref{eq: PL NPIV model}) gives rise to an integral equation whose empirical counterpart is then  ``solved" to obtain estimators of $\phi_0$ and $\beta_0$. This problem, however, is {\itshape ill-posed}, in the sense that the naive solution of such an equation is not  ``stable" and is inconsistent, see Section \ref{sec: Estimation by Landweber Fridman Regularization } for details. To  ``stabilize" such a solution a popular regularization scheme employed in the literature is the Tikhonov regularization, see \cite{darolles_nonparametric_2011} and \cite{carrasco_chapter_2007}. However, from a practical standpoint, estimators based on the Tikhonov regularization require inversions of matrices whose dimension is the sample size, see \cite{centorrino_additive_2017}. Thus, when the sample size is large the Tikhonov estimators will be computationally demanding. Moreover, from a theoretical point of view, the Tikhonov scheme cannot exploit orders of smoothness larger than 2, see \cite{carrasco_chapter_2007}. Differently, the Landweber-Fridman regularization is an iterative method that does not require inverting large matrices and can exploit orders of smoothness larger than 2. Our paper is the first to provide an estimation method for the partly linear IV model entirely based on the Landweber-Fridman scheme. \\
Our third contribution is to obtain the convergence rate for the nonparametric estimator of $\phi_0$ and the asymptotic normality of the parametric estimator of $\beta_0$. This task is technically challenging, due to the Landweber-Fridman regularization scheme. To the best of our knowledge, we are the firsts to establish the asymptotic normality of the parametric estimator of $\beta_0$ based on a Landweber-Fridman scheme. Such an asymptotic normality result is obtained without necessarily relying on the identification of $\phi_0$, see Section \ref{sec: The semiparametric model} for details. \\
Our fourth contribution is to draw a connection between our approach that does not smooth on the IVs and the typical approach that instead does. Typical estimators of $\phi_0$ and $\beta_0$ as in \cite{darolles_nonparametric_2011} or \cite{florens_instrumental_2012} are based on a preliminary estimate of the operator $\mathbb{E}\{\cdot|W\}$. Thus, they smooth on the IVs. We show that our estimator that does not smooth on the IVs coincides with a classical estimator that smooths on the IVs but keeps the bandwidth for
the IVs fixed as the sample size increases, see Section \ref{sec: A fixed bandwidth interpretation} for details. This unveils an interesting feature of the classical estimator. Indeed, when the bandwidth for the IVs is fixed, the estimators of the operators such as $\mathbb{E}\{\cdot|W\}$ will be inconsistent, as their nonparametric bias will not vanish. Thus,  the classical estimator of $\phi_0$ will be based on inconsistent estimators.
By showing consistency of our estimator, we also prove that the classical estimator based on a fixed bandwidth for the IVs will remain {\itshape consistent} although it is based on estimators that are {\itshape inconsistent}. \\
Finally, as a last contribution we propose a data driven method to select the regularization parameter for the estimation of $\phi_0$ and $\beta_0$. In our simulation study, we obtain a satisfying performance of the
estimator of $\beta_0$ both in terms of size coverage and in terms of power. Our simulations also show that our estimator of $\phi_0$ behaves reasonably well. \\

\noindent\textbf{Related literature.} This work is related to the extensive literature on nonparametric and semiparametric IV regressions, see \cite{carrasco_chapter_2007}, \cite{darolles_nonparametric_2011}, \cite{florens_instrumental_2012},
\cite{newey2003instrumental},
\cite{ai2003efficient}, 
\cite{hall_nonparametric_2005}, \cite{horowitz2011applied}, \cite{d2011completeness}, 
\cite{gagliardini2012tikhonov},
\cite{chen2012estimation}, \cite{chen2012estimation},
\cite{johannes2013iterative}, 
\cite{horowitz2014adaptive}, \cite{chetverikov2017nonparametric}, \cite{chen2021robust}, \cite{florens2018nonparametric}, \cite{beyhum2023one}. Estimation procedures for semiparametric IV models based on kernel methods and Tikhonov regularization are provided in \cite{florens_instrumental_2012} and \cite{birke2017semi}, while \cite{ai2003efficient}, \cite{chen2012estimation}, and \cite{chen2021robust} focus on series methods. Such papers estimate the semiparametric IV regressions by making a preliminary smoothing over the IVs or by running preliminary first-stage regressions on the IVs. The main differences between our estimation method and such works are that (i) our method does not smooth on the IVs or does not need running  first-stage nonparametric regressions over the IVs, and (ii) that we do not use the Tikhonov regularization for estimating the IV model. The Landweber-Fridman scheme is also used in \cite{johannes2013iterative} and \cite{florens2018nonparametric} to estimate a fully nonparametric IV model. Our work is different, as (i) we do not smooth on the IVs, (ii) we consider the estimation  of a semiparametric partly linear model involving a parametric component, and (iii) we show the asymptotic normality of the parametric estimator. \\
Finally, our work is related to \cite{escanciano2018simple}, \cite{lavergne_smooth_2013}, and \cite{choi2022generalized} who estimate finite dimensional parameters in endogenous parametric models without smoothing on the IVs.
\cite{lavergne_smooth_2013} and \cite{escanciano2018simple} obtain fixed bandwidth asymptotics for their estimators. 
Our work differs from these papers as in our context, due to the presence of the nonparametric function $\phi_0$, we have to deal with an inverse problem that is ill-posed.\\

\noindent\textbf{Outline.} In Section \ref{sec: The proposed approach and identification} we describe our approach and discuss the identification of $(\beta_0,\phi_0)$. Section \ref{sec: Estimation by Landweber Fridman Regularization } introduces our estimation method for $\beta_0$ and $\phi_0$ based on the Landweber-Fridman regularization. Section \ref{sec: Heuristics and intuition} provides heuristics and intuition about the Landweber-Fridman regularization scheme we use.  Section \ref{sec: A fixed bandwidth interpretation} discusses a fixed bandwidth interpretation of our estimation procedure, showing that our estimator that does not smooth on the IVs coincides with a  typical estimator that smooths on the IVs but keeps the respective bandwidth fixed with the sample size. In Section \ref{sec: Assumptions and Asymptotic Behavior} we state the assumptions, and we obtain the convergence rate for the estimator of $\phi_0$ and the asymptotic normality for the estimator of $\beta_0$. The details about the implementation of our estimators, the Monte Carlo simulations, and the proofs of our results are gathered in a Supplementary Material. 

\section{The Framework, the Proposed Approach, and Identification}\label{sec: The proposed approach and identification}
In Equation (\ref{eq: PL NPIV model}) we can assume without loss of generality that $W$ has a bounded support, as we can always transform each component of $W$ by a bounded one-to-one function.\footnote{In particular, if $f:\mathbb{R}^q\mapsto \mathbb{R}^q$ is a one-to-one mapping, then $\mathbb{E}\{U|W\}=\mathbb{E}\{U|f(W)\}$, as the sigma field generated by $W$ equals the sigma field generated by $f(W)$. } Let us also assume that $\mathbb{E}U^2<\infty$. Our approach is based on Bierens' characterization. So, let $\omega:\mathbb{R}\mapsto \mathbb{C}$ be an analytic non-polynomial function with $\partial^l \omega(0)\neq 0$ for all $l\in\mathbb{N}$, see below for examples.  By \citet[Theorem 2.2]{bierens_econometric_2017}, Equation (\ref{eq: PL NPIV model}) is equivalent to \begin{equation}\label{eq: continuum of unconditional moments}
    \mathbb{E}\{U\,\omega(W^T t)\}=0\text{ for all }t\in\mathcal{T}\, , 
\end{equation}
where $\mathcal{T}\subset \mathbb{R}^q$ is a set containing an arbitrary neighborhood of the origin. Some choices of $\omega$ are $exp(\cdot)$, $cos(\cdot)+sin(\cdot)$, or $exp(\textbf{i}\cdot)$, with $\textbf{i}$ representing the imaginary root. Further choices of $\omega$ can be found in, e.g., \cite{bierens_asymptotic_1997} and \cite{stinchcombe_consistent_1998}. \\
Since $\mathbb{E}\{U|W\}=0$ is equivalent to (\ref{eq: continuum of unconditional moments}), $(\beta_0,\phi_0)$ is identified from Equation (\ref{eq: PL NPIV model}) {\itshape if and only if} it is identified from the following equation
\begin{equation}\label{eq: Bierens equation}
    \mathbb{E}\{Y\, \omega(W^T t)\}= \mathbb{E}\{X^T\beta_0\, \omega(W^T t)\}+\mathbb{E}\{\phi_0(Z)\, \omega(W^T t)\}\, \forall \, t\in\mathcal{T}\, .
\end{equation}
We will express the identification conditions in terms of conditions on the components of Equation (\ref{eq: Bierens equation}). To this end, let us introduce some notation. We assume that $Y$ and $X$ are square integrable, $\sup_{t\in\mathcal{T}}|\omega(W^T t)|$ is bounded, and we let $\mu$ be a positive finite measure supported on $\mathcal{T}$. For example, $\mu$ could be set to the (truncated) normal distribution supported on $\mathcal{T}$. Also, let $L^2_{\mu}(\mathcal{T})$ be the space of function defined on $\mathcal{T}$ that are square integrable with respect to $\mu$. We define
\begin{equation}\label{eq: definition of s}
    s(t):=\mathbb{E}\{Y\,\omega(W^T t)\} \, ,\,  s\in L^2_\mu (\mathcal{T})
\end{equation}
and for all $\beta\in \mathbb{R}^\kappa$ and $t\in\mathcal{T}$
\begin{equation}\label{eq: definition of AX}
    (A_X \beta)(t) :=\mathbb{E}\{X^T\beta\, \omega(W^T t)\}\, . 
\end{equation}
Since $\sup_{t\in\mathcal{T}}|\omega (W^T t)|<\infty$ and each component of $X$ has finite second moments, the expectation on the right hand side of the previous equation is well defined and 
\begin{equation*}
    A_X:\mathbb{R}^\kappa \mapsto L^2_\mu(\mathcal{T})\, .
\end{equation*}
Let $\pi$ be a density function that is strictly positive on the the support of $Z$, and let $L^2_\pi (\mathbb{R}^p)$ be the space of functions that are square integrable with respect to $\pi$. The measure $\pi$ is introduced for technical reasons. In particular, from a technical point of view it would be ideal to work with the space $L^2(Z)$ of square integrable functions with respect to $Z$. However, since we do not know the distribution of $Z$, we cannot directly use the space $L^2(Z)$. Thus, we replace $L^2(Z)$ with the known space $L^2_\pi(\mathbb{R}^p)$ and work with the latter. By denoting with $f_Z$ the density of $Z$, we assume that $f_Z/\pi \in L^2_\pi(\mathbb{R}^p)$. Then, 
for any $\phi\in L^2_\pi(\mathbb{R}^p)$
\begin{align}\label{eq: def of A_Z}
    (A_Z \phi)(t):=\mathbb{E}\{\phi(Z)\, \omega(W^T t)\}=\int_{ }\phi(z)\,\mathbb{E}\{\omega(W^T t)|Z=z\}f_Z(z) dz \, ,
\end{align}
where in the second equality we have used the law of iterated expectations. Notice that since $f_Z/\pi \in L^2_\pi (\mathbb{R}^p)$ and $\sup_{t\in\mathcal{T}}|\omega(W^T t)|$ is bounded, the integral on the right hand side of the previous equation is well defined and\footnote{Since $\sup_{t\in\mathcal{T}}|\omega(W^T t)|<C$ and $(A_Z\phi)(t) =\int \phi(z)\mathbb{E}\{\omega(W^T t)|Z=z\} [f(z)/\pi(z)] \, \pi(dz)$, by the Cauchy-Schwartz inequality  $|(A_Z \phi) (t)|^2\leq C^2 \int |\phi(z)|^2 \pi(dz)\, \int |f(z)/\pi(z)|^2 \pi(dz)$. So, the condition $f_Z/\pi \in L^2_\pi(\mathbb{R}^p)$ ensures that $A_Z\phi \in L^2_\mu(\mathcal{T})$ for any $\phi\in L^2_\pi(\mathbb{R}^p)$. } 
\begin{equation*}
    A_Z:L^2_\pi(\mathbb{R}^p)\mapsto L^2_\mu(\mathcal{T})\, .
\end{equation*}
We complete the presentation of the framework by introducing $A_X^*$ and $A_Z^*$, the Hilbert adjoints of $A_X$ and $A_Z$. These will be needed to set up the regularization scheme and the estimation procedure in the next section. Let us define  
\begin{equation}\label{eq: def of k}
    k(z,t):=\mathbb{E}\{\omega(W^T t)|Z=z\}f_Z(z)\, .
\end{equation}
From Equation (\ref{eq: def of A_Z}), $A_Z$ can be rewritten as an integral operator that depends on the above kernel, so 
\begin{equation}\label{eq: AZ as an integral operator}
     (A_Z\phi)(t)=\int_{ }\phi(z) k(z,t) dz \, .
\end{equation}
Let us denote with $\left<\cdot,\cdot\right>$ the inner product on $L^2_\pi(\mathbb{R}^p)$, so that $\left<\phi,\varphi\right>:=\int \phi(z) \overline{\varphi}(z) \pi (dz)$ for all $\phi,\varphi \in L^2_\pi(\mathbb{R}^p)$, where $\overline{\varphi}(z)$ denotes the complex conjugate of $\varphi(z)$. We denote with $\|\cdot\|$ the norm induced by the inner product $\left<\cdot,\cdot\right>$. The inner product and the norm on $L^2_\mu(\mathcal T)$ are similarly defined and we will denote them also by $\left<\cdot,\cdot\right>$ and $\|\cdot\|$. At each time, the specific space the inner product and the norm refer to will be clear from their arguments. Notice that the operator $A_Z$ is bounded, in the sense that there exists a constant $C$ such that $\|A_Z\phi\|\leq C \|\phi\|$ for all $\phi\in L^2_\pi(\mathbb{R}^p)$.\footnote{To see that $A_Z$ is a bounded operator, notice that $|(A_Z\phi)(t)|^2 \leq $ $C \int |\phi(z)|^2 \pi(dz)\, \int |f(z)/\pi(z)|^2 \pi(dz)$ for a fixed constant $C$, so we have $\|A_Z\phi\|^2=\int|(A_Z \phi)(t)|^2\mu(dt)$ $\leq C^* \|\phi\|^2 $ for a constant $C^*$. } When an operator is bounded it always admits a Hilbert adjoint, see \cite[Theorem 4.9]{kress_linear_2012}. The Hilbert adjoint of $A_Z$ is the operator $A_Z^*:L^2_\mu(\mathcal T)\mapsto L^2_\pi(\mathbb{R}^p)$ such that $\left< A_Z \phi, \psi \right>=\left< \phi, A_Z^* \psi \right>$  for all $\phi\in L^2_\pi(\mathbb{R}^p) $ and $\psi \in L^2_\mu(\mathcal T) $, see \citet[Chapter 4]{kress_linear_2012}. Given the structure of $A_Z$, by a direct computation we find that its Hilbert adjoint is
\begin{equation}\label{eq: definition of A_Z^*}
    (A_Z^*\psi)(z)=\int_{ }\psi(t) \frac{\overline{k}(z,t)}{\pi(z)} \mu(dt)\, . 
\end{equation}
where $\overline{k}$ denotes the complex conjugate of $k$ and\footnote{Notice that since $\sup_{t\in\mathcal{T}}|\omega(W^T t)|\leq C$, the ratio $\overline{k}(z,t)/\pi(z)$ is bounded in absolute value by  $C\,f_Z(z)/\pi(z)$. Thus, by the Cauchy-Schwartz inequality $|(A_Z^*\psi)(z)|^2\leq \int |\psi(t)|^2\mu(dt)\int |k(z,t)/\pi(z)|^2\mu(dt)$ $\leq \int |\psi(t)|^2\mu(dt)\, C^2\, \mu(\mathcal{T})\, |f_Z(z)/\pi(z)|^2$ with $f_Z/\pi\in L^2_\pi(\mathbb{R}^p)$. This gives $A_Z^*\psi\in L^2_\pi(\mathbb{R}^p)$.  }  
\begin{equation*}
     A_Z^*:L^2_\mu(\mathcal T)\mapsto L^2_\pi(\mathbb{R}^p)\, .
\end{equation*}
Similarly, $A_X$ defined in Equation (\ref{eq: definition of AX}) is also a bounded operator, in the sense that for a fixed constant $C$ we have $\|A_X\beta\|\leq C \|\beta\|$ for all $\beta\in\mathbb{R}^\kappa$, where $\|\beta\|$ denotes the Euclidean norm of $\beta$.\footnote{$|(A_X\beta)(t)|^2= |\mathbb{E}\{X^T\beta\,\omega(W^T t)\}|^2$ $\leq C^2 \mathbb{E}\{\|X\|^2\}\|\beta\|^2$, thus $\|A_X\beta\|^2=\int |(A_X\beta)(t)|^2\mu(dt)$ $\leq C^* \|\beta\|^2$ for a constant $C^*$, so that $A_X$ is a bounded operator.} Thus, $A_X$ will also admit a Hilbert adjoint. By a direct computation, we find that the Hilbert adjoint of $A_X$ is 
\begin{equation}\label{eq: definition of A_X^*}
    A_X^*g=\mathbb{E}\, X\int_{ }\overline{\omega}(W^T t)g(t)\mu(dt)\, 
\end{equation}
with 
\begin{equation*}
    A_X^*:L^2_\mu(\mathcal{T})\mapsto \mathbb{R}^\kappa\, .
\end{equation*}
Let us denote with $\mathcal{R}(A_X)$ and $\mathcal{R}(A_Z)$ the ranges of $A_X$ and $A_Z$.\footnote{Formally, $\mathcal{R}(A_Z):=\{b\in L^2_\mu(\mathcal{T})\,:\, b=A_Z \phi\text{ for some }\phi\in L^2_\pi(\mathbb{R}^p)\}$ and $\mathcal{R}(A_X):=\{b\in L^2_\mu(\mathcal{T})\,:\, b=A_X \beta \text{ for some }\beta\in \mathbb{R}^\kappa\}$. } We denote with $P_Z:L^2_\mu(\mathcal T)\mapsto L^2_\mu(\mathcal T) $ the projection operator onto $\overline{\mathcal{R}(A_Z)}$, the closure of $\mathcal{R}(A_Z)$. We can now express the identification conditions on $(\beta_0,\phi_0)$ in terms of conditions on $A_X$ and $A_Z$.




\begin{thm}\label{prop: identification}
    Assume that $Y$ and each component of $X$ have finite second moments, $\sup_{t\in\mathcal{T}}|\omega(W^T t)|$ is bounded, and $f_Z/\pi\in L^2_\pi(\mathbb{R}^p)$. Let us introduce the following two conditions
     \begin{enumerate}
\item $A_X$ is injective\footnote{The operator $A_X$ is injective if and only if $A_X\beta=0\Leftrightarrow \beta=0$. The same holds for $A_Z$.}
\item $\mathcal{R}(A_X)\cap \overline{\mathcal{R}(A_Z)}=\{0\}$.
\end{enumerate}
Then, \\

\noindent \textbf{(i)} Conditions 1 and 2 $\Leftrightarrow$ $\Sigma:=A_X^* (I-P_Z)A_X$ is injective $\Leftrightarrow$ $\beta_0$ is identified as 
\begin{equation}\label{eq: beta0 identification expression}
\beta_0=\Sigma^{-1} A_X^*(I-P_Z)s    
\end{equation}

\noindent \textbf{(ii)} If $\beta_0$ is identified and $A_Z$ is injective, then $\phi_0$ is identified as 
\begin{equation}\label{eq: phi0 identification expression}
\phi_0=A_Z^{-1}(s-A_X\beta_0)\, .    
\end{equation}

\end{thm}

\begin{proof}
    \textbf{(i)} We start by proving the first equivalence. Assume that conditions 1 and 2 hold. Since $P_Z$ is a projection operator, $P_Z^*=P_Z$ and $P_Z P_Z=P_Z$ so that $(I-P_Z)^*(I-P_Z)=(I-P_Z)$. Thus,  $\Sigma \beta=0$ implies that $0=\left<A_X^* (I-P_Z) A_X\beta,\beta\right>=\left<A_X^* (I-P_Z)^*(I-P_Z) A_X\beta,\beta\right>$$=\left<(I-P_Z)A_X\beta,(I-P_Z)A_X\beta\right>=\|(I-P_Z)A_X\beta\|^2$. 
    This in turn implies that $(I-P_Z)A_X\beta=0$. Hence, $A_X\beta=P_Z A_X\beta\in \overline{\mathcal{R}(A_Z)}$ and  $A_X\beta\in\mathcal{R}(A_X)\cap \overline{\mathcal{R}(A_Z)}$. Since $\mathcal{R}(A_X)\cap \overline{\mathcal{R}(A_Z)}=\{0\}$ by condition 2, we get $A_X\beta=0$. This implies $\beta=0$, by injectivity of $A_X$ in condition 1. We have therefore proved that $\Sigma\beta=0$ implies $\beta=0$ and hence that $\Sigma$ is injective under conditions 1 and 2. Let us  now prove that injectivity of $\Sigma$ implies conditions 1 and 2. First notice that $\Sigma$ can be injective only if condition 1 holds. In fact, if condition 1 did not hold and $A_X$ was not injective, there would exist $\beta\neq 0$ with $A_X\beta=0$, which in turn would imply that $\Sigma\beta=0$ with $\beta\neq 0$ and hence the non-injectivity of $\Sigma$. To show that the injectivity of $\Sigma$ also implies condition 2, let us pick $\widetilde s \in \mathcal{R}(A_X)\cap \overline{\mathcal{R}(A_Z)}$. Then, $\widetilde s=A_X \widetilde{\beta}$ for some $\widetilde{\beta}\in\mathbb{R}^\kappa$ and $A_X \widetilde{\beta}=P_Z A_X \widetilde \beta$. Hence,   $(I-P_Z)A_X\widetilde{\beta}=0$ which implies that $\Sigma \widetilde \beta=A_X^* (I-P_Z) A_X \widetilde \beta=0$. This implies $\widetilde \beta=0$ by injectivity of $\Sigma$, and hence $\widetilde s=0$. Thus, $\mathcal{R}(A_X)\cap \overline{\mathcal{R}(A_Z)}=\{0\}$. Hence, we have proved that injectivity of $\Sigma$ implies condition 2.\\
    We now show that injectivity of $\Sigma$ implies identification of $\beta_0$. Given the definitions of $A_Z$, $A_X$, and $s$, Equation (\ref{eq: Bierens equation}) can be written as 
    \begin{equation}\label{eq: Bierens equation in terms of AX and AZ}
        s=A_X \beta_0 + A_Z \phi_0\, .
    \end{equation}
    Since $(I-P_Z) A_Z=0$, by applying the operator $(I-P_Z)$ to both sides of the previous equation we get $(I-P_Z)s=(I-P_Z)A_X \beta_0$, and hence $A_X^*(I-P_Z)s=\Sigma \beta_0$. By injectivity of $\Sigma$, we finally obtain $\beta_0=\Sigma^{-1}A_X^*(I-P_Z) s$. \\

    \textbf{(ii)} When $A_Z$ is injective, $A_Z^{-1}$ exists. So,  $\phi_0=A_Z^{-1}(s-A_X\beta_0)$ is obtained from (\ref{eq: Bierens equation in terms of AX and AZ}). 
\end{proof}
Let us now comment on the conditions of Theorem \ref{prop: identification}. Injectivity of $A_X$ in Condition 1 is equivalent to the linear independence of $\mathbb{E}\{X|W\}$. In fact, by \citet[Theorem 2.2]{bierens_econometric_2017} $(A_X \beta)=\mathbb{E}\{X^T\beta\, \omega(W^T \cdot)\}=0$ if and only if $\mathbb{E}\{X^T|W\}\beta=0$. This latter equality will be equivalent to $\beta=0$ if and only if $\mathbb{E}\{X|W\}$ are linearly independent. 
Hence, the linear independence of $\mathbb{E}\{X|W\}$ is necessary and sufficient for the injectivity of $A_X$. 
Condition 2 requires that the range of $A_X$ and (the closure of) the range of $A_Z$ must have in common only the identically zero function.\footnote{Notice that since $A_X$ is defined on $\mathbb{R}^\kappa$, its range is a linear finite dimensional space. Since linear finite dimensional spaces are closed, see \citet[Theorem 2.4-2]{kreyszig1991introductory}, the range of $A_X$ is also closed.} This means that these two ranges must be well ``separated". Indeed, if this did not happen, injectivity of $\Sigma$ would not hold and hence $\beta_0$ could not be identified as in Theorem \ref{prop: identification}. As stated in Theorem \ref{prop: identification}, Conditions 1 and 2 are equivalent to a full-rank condition of the matrix $\Sigma$ which is testable. Accordingly, when $\mathbb{E}\{X|W\}$ are linearly independent (and hence Condition 1 holds), Condition 2 can be tested by checking that $\Sigma$ is full rank. Injectivity of $A_Z$ is equivalent to the {\itshape completeness} condition introduced in \cite{newey2003instrumental}. In particular, by  \cite[Theorem 2.2]{bierens_econometric_2017}  $A_Z\phi=\mathbb{E}\{\phi(Z)\omega(W^T \cdot)\}=0$ if and only if $\mathbb{E}\{\phi(Z)|W\}=0$. This last equality will be equivalent to $\phi=0$ if and only if the distribution of $Z$ conditional on $W$ is complete, see \cite{newey2003instrumental}.\footnote{Formally, the distribution of $Z$ conditional on $W$ is complete if $\mathbb{E}\{g(Z)|W\}=0\Rightarrow\,g(Z)=0$ a.s. for any function $g\in L^2(Z)$.} Hence, such a completeness condition is equivalent to the injectivity of $A_Z$. Completeness is a common assumption in nonparametric IV models, see, e.g., \cite{darolles_nonparametric_2011} and \cite{newey2003instrumental}.  \\
According to Theorem \ref{prop: identification}, identification of $\beta_0$ does not rely on the injectivity of $A_Z$ and hence on the completeness condition. 
Thus, estimation of $\beta_0$ can be obtained without injectivity of $A_Z$ and hence without necessarily relying on the identification of $\phi_0$. See Section \ref{sec: The semiparametric model} for details.

For clarification purposes, it might be useful to link the identification of the partly linear IV model to the more classical order and rank conditions typically used for the parametric linear IV models. \\

\textbf{Example}. Let $S:=(X^T,Z^T)^T$ and $(S^T,W^T)^T\sim\mathcal{N}(0,\Sigma)$. We define  $\Sigma_{S W}:=\text{Cov}(S,W)$ and $\Sigma_{S S}:=\text{Var}(S)$. 
In this example we show that if $\Sigma_{S S}$ is full rank and $\text{Rank}(\Sigma_{S S})=\text{Rank}(\Sigma_{S W})$, then $(\beta_0,\phi_0)$ are identified. Thus, in this simple case, the classical order and rank conditions typical of linear IV models ensure the identification in the semiparametric partly linear model. To show this, from \cite{florens1990elements} the condition $\text{Rank}(\Sigma_{S S})=\text{Rank}(\Sigma_{S W})$ implies that $S$ is strongly identifiable by $W$, in the sense that $\mathbb{E}\{g(S)|W\}=0\Rightarrow g(S)=0$ for any square integrable function $g$. Thus,  $\mathbb{E}\{X^T\beta+\phi(Z)|W\}=0\Rightarrow X^T\beta+\phi(Z)=0$. Given the joint normality of $S=(X^T,Z^T)^T$, $X^T\beta=-\phi(Z)$ implies that $\phi=0$ and $\beta^T X=0$. Since $\Sigma_{S S}$ is full rank, the components of $X$ are linearly independent and  $\beta^T X=0\Rightarrow \beta=0$. Hence, $\mathbb{E}\{X^T\beta+\phi(Z)|W\}=0$ implies $\beta=0$ and $\phi=0$. Since from \citet[Theorem 2.2]{bierens_econometric_2017} $\mathbb{E}\{X^T\beta+\phi(Z)|W\}=0$ is equivalent to $\mathbb{E}\{[X^T\beta+\phi(Z)]\omega(W^T\cdot)\}=0$, we have that $\mathbb{E}\{[X^T\beta+\phi(Z)]\omega(W^T\cdot)\}=0$ implies $\beta=0$ and $\phi=0$. Thus, $\mathbb{E}\{[Y-X^T\beta-\phi(Z)]\omega(W^T\cdot)\}=0$ can only be satisfied at $(\beta_0,\phi_0)$, so that $\beta_0$ and $\phi_0$ are identified from such an equation. It is easy to check that when $\mathbb{E}\{[X^T\beta+\phi(Z)]\omega(W^T\cdot)\}=0$ implies $\beta=0$ and $\phi=0$, we have that $A_Z$ and $A_X$ are injective and $\mathcal{R}(A_Z)\cap \mathcal{R}(A_X)=\{0\}$.\\

By Theorem \ref{prop: identification}, Assumption \ref{ass: identification} below ensures that $\beta_0$ and $\phi_0$ are identified. \\
\begin{hp}\label{ass: identification}
(a) $A_X$ is injective and $\mathcal{R}(A_X)\cap \overline{\mathcal{R}(A_Z)}=\{0\}$; (b)  $A_Z$ is injective. 
\end{hp}
Before moving to the estimation of the model, we remark that our approach allows for common components between the regressors $(X,Z)$ and the instruments $W$. Thus, there can be exogenous control variables that enter the partly linear regression and the IVs. This feature is not shared by the classical approach that smooths on the IVs. See Section \ref{sec: A fixed bandwidth interpretation} for details.

\section{Estimation by Landweber-Fridman Regularization}\label{sec: Estimation by Landweber Fridman Regularization }
To estimate $\phi_0$ and $\beta_0$, we will need to rely on regularization techniques, as it is common in semiparametric IV models. The following section provides heuristics and intuition about the regularization techniques in our context. 



\subsection{Heuristics and Intuition}\label{sec: Heuristics and intuition}
To introduce smoothly the estimation method, let us first assume to have consistent estimators $\widehat{s}$, $\widehat{A}_X$, and $\widehat \beta$ of their population counterparts. Also, let us assume that $A_Z$ is known. Under Assumption  \ref{ass: identification}\textcolor{red}{(b)}, $A_Z$ is one-to-one so $\phi_0=A_Z^{-1}(s-A_X \beta_0)$. From this expression, it would be tempting to estimate $\phi_0$ by $\widehat{\phi}=A_Z^{-1}(\widehat{s}-\widehat{A}_X \widehat \beta)$: since $\widehat{s}$, $\widehat{A}_X$, and $\widehat{\beta}$ are consistent, $\| \widehat{s}-\widehat{A}_X\widehat \beta\|=o_P(1)$, and we would expect that  $\|\widehat{\phi}-\phi_0\|=o_P(1)$ by a ``continuous mapping theorem". However, in this context such an argument will not hold. In fact, although $A_Z$ is one-to-one, its inverse $A_Z^{-1}$ is not continuous, so the convergence of
 $\widehat{s}-\widehat{A}_X\widehat \beta$ towards $s-A_X \beta_0$ will not imply the consistency of $\widehat{\phi}$. This is because of  the well known {\itshape ill-posedness} of the inverse problem: since $A_Z$ is an integral operator, see Equation (\ref{eq: AZ as an integral operator}), its inverse $A_Z^{-1}$ is not continuous.\footnote{\label{footnote: about the definition of kernel operator, compactness, and non continuous inverse} An operator $\mathcal{K}:L^2_\pi(\mathbb{R}^p)\mapsto L^2_\mu(\mathcal{T})$ is an integral/kernel operator if there exists a function $b:\mathbb{R}^p\times \mathcal{T}\mapsto \mathbb{C}$ (with $\int|b(z,t)|^2 \pi(z)\,\mu(t) \,dz\,dt<\infty$) such that $(\mathcal{K}\phi)(t)=\int \phi(z)b(z,t)\pi(d z)$ for all $\phi\in L^2_\pi(\mathbb{R}^p)$, see \cite[Example 2.2]{carrasco_chapter_2007}. From Equation (\ref{eq: AZ as an integral operator}), $A_Z$ is an integral/kernel operator. This implies that it is compact, see \cite[Theorems 2.32 and 2.34]{carrasco_chapter_2007}. $A_Z$ is a compact operator in the sense that for any bounded sequence $(\phi_j)_j$ in $L^2_\pi(\mathbb{R}^p)$ the sequence $(A_Z\phi_j)_j$ admits a convergent subsequence in $L^2_\mu(\mathcal{T})$, see \cite[Theorem 2.13]{kress_linear_2012}. Compact operators that are injective do not have a  continuous inverse. See Theorems 2.16 and 2.20 in \cite{kress_linear_2012}. } \\
 To deal with the lack of continuity of $A_Z^{-1}$, we replace $A_Z^{-1}$ by a regularization scheme.  Formally, a regularization scheme is a sequence of operators $R_m(A_Z):L^2_\mu (\mathcal T)\mapsto L^2_\pi (\mathbb{R}^p)$ indexed by $m\in \mathbb{N}$ such that (i) each $R_m(A_Z)$ is continuous and (ii) for each $b\in \mathcal{R}(A_Z)$ it holds that $R_m(A_Z)b\rightarrow A_Z ^{-1}b$ as $m\rightarrow \infty$. So, for a large $m$, $R_m(A_Z)$ will be close enough to $A_Z^{-1}$ (in a pointwise sense) and at the same time will be continuous. $m$ is called the {\itshape regularization parameter}. 
A popular scheme used in the literature is the Tikhonov regularization, see e.g. \cite{darolles_nonparametric_2011}.\footnote{The Tikhonov regularization scheme is $(I m^{-1}+A^*_{Z}A_Z)^{-1}A^*_Z $, where $I$ is the identity operator.} In this paper we will instead adopt
the Landweber-Fridman regularization. To the best of our knowledge, we are the first to employ the Landweber-Fridman scheme to estimate a {\itshape partly linear} IV regression. There are two reasons motivating our choice. First, the Tikhonov regularization requires the inversion of matrices whose order is the sample size, see \cite{centorrino_additive_2017}. So, if the sample size is large the Tikhonov regularized estimator will be computationally intense. Second, the Tikhonov regularization scheme cannot exploit  ``smoothness orders" of $\phi_0$ larger than 2, see \cite{carrasco_chapter_2007}. Differently, the Landweber-Fridman scheme is an iterative method that does not require the inversion of large matrices and can exploit smoothness orders larger than 2. To describe the Landweber-Fridman regularization, let us recall that $A_Z^*$ is the Hilbert adjoint of $A_Z$ (see the previous section) and let us denote with $\|A_Z\|_{op}$ the operator norm of $A_Z$. Formally, 
\begin{equation*}
\|A_Z\|_{op}:=\sup_{\varphi\in L^2_\pi (\mathbb{R}^p)\,,\,\|\varphi\|=1}\|A_Z \varphi\|\, ,
\end{equation*}
where $\|A_Z\varphi\|^2=\int|(A_Z\varphi)(t)|^2\mu(dt)$. 
Let $a$ be a fixed constant such that $0<a<1/\|A_Z\|^2_{op}$. Then, $R_m(A_Z)(\widehat s - \widehat A _X \widehat \beta)$ is computed according to the following iterations:
\begin{enumerate}
    \item[1] initialize with $\varphi_0=a A^*_Z (\widehat{s}-\widehat{A}_X \widehat{\beta})$
    \item[2] for $l=1,\ldots,m$ compute $\varphi_l=(I-a A^*_Z A_Z) \varphi_{l-1}+\varphi_0$
    \item[3] set $R_m(A_Z)(\widehat{s}-\widehat{A}_X \widehat{\beta})=\varphi_m$ .
\end{enumerate}
If $A_Z$ was known, the final estimator of $\phi_0$ would be $\widehat{\phi}=R_m(A_Z)(\widehat{s}-\widehat{A}_X \widehat{\beta})$. Notice that such an estimator is obtained from Equation (\ref{eq: phi0 identification expression}), where we replace $A_Z^{-1}$ (which is not continuous) with $R_m(A_Z)$ (that is continuous). Intuitively, given the consistency of $\widehat s$, $\widehat A _X$, and $\widehat \beta$, and the continuity of $R_m(A_Z)$, a continuous mapping theorem will hold for  $\widehat{\phi}=R_m(A_Z)(\widehat{s}-\widehat{A}_X \widehat{\beta})$. 
The Landweber Fridman scheme can also be written 
 as 
\begin{equation}\label{eq: def of R_m(A_Z) for LF}
    R_m(A_Z)=a\sum_{l=0}^m(I-a A^*_Z A_Z)^l A_Z^* \, ,
\end{equation}
 see \cite[Thereom 15.27 and Equation 15.45]{kress_linear_2012}. In practice  $A_Z$ is unknown, so it will be replaced by a consistent estimator $\widehat{A}_Z$ that we define in the next section. 

\subsection{Definition of $\widehat{\phi}$ and $\widehat{\beta}$}
In this section we introduce in detail the estimation procedure for $\beta_0$ and $\phi_0$. 
Since the estimation of $(\beta_0,\phi_0)$ is based on Equations (\ref{eq: beta0 identification expression}) and (\ref{eq: phi0 identification expression}), we will need to estimate $s$, $A_X$, $A_Z$, and $P_Z$. From Equation (\ref{eq: definition of s}) we estimate $s$ by a sample average of stochastic processes 
\begin{equation}\label{eq: def of s hat}
    \widehat{s}(t):=\mathbb{E}_n Y \omega(W^T t)\, 
\end{equation}
where $\mathbb{E}_n$ denotes the empirical mean operator.\footnote{Formally, $\mathbb{E}_n g(Y,X,Z,W):=(1/n)\sum_{i=1}^n g(Y_i,X_i,Z_i,W_i)$ for any function $g$.} Similarly, from Equation (\ref{eq: definition of AX}) $A_X$ is estimated as 
\begin{equation}
(\widehat{A}_X \beta)(t):=\mathbb{E}_n\, X^T\, \beta \, \omega(W^T t) \, , \,\, \widehat A_X:\mathbb{R}^\kappa\mapsto L^2_\mu(\mathcal{T})\, .    
\end{equation}
$A_X^*$ is estimated by taking the Hilbert adjoint of $\widehat{A}_X$. So, by a direct computation we get 
\begin{equation}
    \widehat{A}_X^* g := \mathbb{E}_n X \int_{ }\overline{\omega}(W^T t) g(t) \mu (dt)\, , \, \widehat A _X:L_\mu^2(\mathcal{T})\mapsto \mathbb{R}^\kappa\, .
\end{equation}
Notice that $\widehat{A}^*_X$ is exactly the empirical counterpart of $A_X^*$ in (\ref{eq: definition of A_X^*}). From the expressions just introduced, $\widehat s$, $\widehat A_X$, and $\widehat A_X^*$ are all estimated at parametric rates, see Lemma \textcolor{red} {C.2} of the Supplementary Material 
for details. \\
To estimate $A_Z$ and its adjoint, we will use a kernel method. So, let $K$ be a symmetric kernel and $h$ be a bandwidth converging to zero. From Equation (\ref{eq: def of k}) we estimate $k$ by 
\begin{equation}\label{eq: def of k hat}
    \widehat{k}(z,t):=\frac{1}{n h^p}\sum_{i=1}^n \omega(W_i^T t ) K\left( \frac{z-Z_i}{h} \right) \, . 
\end{equation}
Let us assume that $K((\cdot-Z_i)/h) / \pi \in L^2_\pi(\mathbb{R}^p)$. Then, the estimator of $A_Z$ can be obtained as 
\begin{equation}\label{eq: def of A hat_Z}
     (\widehat{A}_Z \phi)(t) = \int_{ }\phi(z) \widehat{k}(z,t) dz\, ,\, \widehat{A}_Z:L^2_\pi(\mathbb{R}^p)\mapsto L^2_\mu(\mathcal T)\, .
\end{equation}
Notice that estimating $A_Z$ does not require selecting a smoothing parameter (a bandwidth) for the IVs but only for $Z$. By a direct computation, its Hilbert adjoint is 
\begin{equation}\label{eq: def of A hat Z^*}
    (\widehat{A}_Z^* \psi)(z) = \int_{ }\psi(t) \frac{\overline{\widehat{k}}(z,t)}{\pi(z)} \mu(dt)\, , \, \widehat{A}^*_Z:L^2_\mu(\mathcal T) \mapsto L^2_\pi(\mathbb{R}^p)\, .
\end{equation}
We notice that $\widehat A_Z^*$ is exactly the empirical counterpart of $A_Z^*$ in (\ref{eq: definition of A_Z^*}).
Let us now discuss the estimation of $P_Z$, the projection operator onto $\overline{\mathcal{R}(A_Z)}$. Such an operator could be estimated by a Tikhonov regularization, similarly as in \cite{florens_instrumental_2012}. Differently, for the reasons highlighted in the previous section, we will here use a Landweber-Fridman scheme for estimating $P_Z$. This is a new estimation approach for such a projection operator. 
In Proposition \textcolor{red}{C.1} of the Supplementary Material, 
 we show that 
  \begin{equation*}
     A_Z R_m(A_Z) g \rightarrow P_Z g\quad  \forall \quad  g\in L^2_\mu(\mathcal T)\, .
 \end{equation*}
So, if $A_Z$ was known, the projection operator $P_Z$ could be estimated by $A_Z R_m(A_Z)$. 
 Hence, using the expression of $R_m(A_Z)$ in (\ref{eq: def of R_m(A_Z) for LF}), we could estimate $P_Z$ by the Landweber-Fridman scheme as 
 \begin{equation*}
    P_Z^m:=A_Z R_m(A_Z)=A_Z a \sum_{l=0}^m (I-a A_Z^* A_Z)^l A_Z^* \, .
 \end{equation*}
 In practice, $A_Z$ is unknown, so we replace it by its estimator $\widehat A _Z$ in (\ref{eq: def of A hat_Z}) and estimate $P_Z$ as  
 \begin{equation}\label{eq: definition of Phat}
    \widehat{P}_Z^m := \widehat{A}_Z R_m(\widehat{A}_Z)=a \widehat{A}_Z  \sum_{l=0}^m (I-a \widehat{A}^*_Z \widehat{A}_Z)^l \widehat{A}^*_Z \,  .
 \end{equation}
 Then, using the expression in (\ref{eq: beta0 identification expression}), we estimate $\beta_0$ as 
 \begin{align}\label{eq: definition of betahat}
    \widehat{\beta}:= \widehat{\Sigma}^{-1}\widehat{A}^*_X (I-\widehat{P}^m_Z)\widehat{s}\, &\qquad\text{ with} \qquad  \widehat{\Sigma}:= \widehat{A}^*_X (I-\widehat{P}_Z^m)\widehat{A}_X\, . 
 \end{align}
 To obtain an estimate of $\phi_0$ we use the expression in (\ref{eq: phi0 identification expression}). The unknown $s$, $\beta_0$, and $A_X$  are replaced by their estimators introduced earlier. As discussed in the previous section, since $A_Z^{-1}$ is not continuous we would replace $A_Z^{-1}$ by $R_m(A_Z)$ if $A_Z$ was known. Since $A_Z$ is unknown, instead of replacing  $A_Z^{-1}$ by  $R_m(A_Z)$ we replace it by $R_m(\widehat A_Z)$. Then, we estimate $\phi_0$ by 
 \begin{equation}\label{eq: definition of phihat}
    \widehat{\phi}= R_m(\widehat A_Z)(\widehat s - \widehat A_X \widehat \beta)=a  \sum_{l=0}^m (I-a \widehat{A}^*_Z \widehat{A}_Z)^l \widehat{A}^*_Z (\widehat{s}-\widehat{A}_X \widehat{\beta})\, ,
 \end{equation}
 where in the second equality we have used (\ref{eq: def of R_m(A_Z) for LF}). 
$\widehat \phi$ can be computed by the iterations reported in Section \ref{sec: Heuristics and intuition}, where $A_Z$ is replaced by $\widehat{A}_Z$. In practice, as we detail in Section \textcolor{blue}{A} of the Supplementary Material, 
 to compute $\widehat \beta$ and $\widehat \phi$ we do not need to compute $\widehat s$ and $\widehat A_Z$ at every value of $t\in\mathcal{T}$, and we do not need to compute $\widehat A_Z^*$ at all value of $z$. \\
We close this section by discussing briefly an alternative approach that we could have used to estimate $(\beta_0,\phi_0)$. Instead of building estimators of $(\beta_0,\phi_0)$ based on (\ref{eq: beta0 identification expression}) and (\ref{eq: phi0 identification expression}), we could have started directly from Equation (\ref{eq: Bierens equation in terms of AX and AZ}). Such an equation is featured by the operator $(\beta,\phi)\mapsto D(\beta,\phi):= A_X \beta + A_Z \phi$. This operator is defined on the Hilbert space $(\mathbb{R}^\kappa \times L^2_\pi(\mathbb{R}^p),\left<\cdot,\cdot\right>)$, with $\left<(\beta_1,\phi_1),(\beta_2,\phi_2)\right>=\beta_1^T \beta_2 + \left<\phi_1,\phi_2\right>$, and takes values in $L^2_\mu(\mathcal{T})$. Now, the injectivity of $A_Z$ and $A_X$ together with Condition 2 of Theorem \ref{prop: identification} imply that $D$ is injective. However, for the same arguments as in Section \ref{sec: Heuristics and intuition} its inverse $D^{-1}$ will not be continuous. 
So, we could estimate the couple $(\beta_0,\phi_0)$ by regularizing $D$. This, however, would imply an unnecessary regularization of the part of $D$ defined on the finite dimensional space $\mathbb{R}^\kappa$, i.e. $A_X$. Such a regularization would not be necessary, as the ill-posedness of the inverse problem in (\ref{eq: Bierens equation in terms of AX and AZ}) only stems from the fact that $A_Z^{-1}$ is not continuous. Thus, it is only with respect to $A_Z$ that we need to regularize. 
Accordingly, to avoid unnecessary regularizations, we construct our estimator of $\beta_0$ and $\phi_0$ by using the expressions in (\ref{eq: beta0 identification expression}) and (\ref{eq: phi0 identification expression}).

\section{A Fix Bandwidth Interpretation}\label{sec: A fixed bandwidth interpretation}
In this section, we show an interesting connection between our approach that does not smooth on the IVs and the typical approach which estimates $\phi_0$ by smoothing on the IVs. To simplify the exposition, let us consider the fully nonparametric model (where $\beta_0\equiv 0$),
\begin{equation*}
    Y=\phi_0(Z)+U\text{ with }\mathbb{E}\{U|W\}=0\, .
\end{equation*}
The integral equation associated to such a model is
\begin{equation*}
    s=A_Z \phi_0\, ,
\end{equation*}
where $s$ is defined in Equation  (\ref{eq: definition of s}) and $A_Z$ in (\ref{eq: def of A_Z}). By denoting with $\textbf{i}$ the imaginary root, we set $\omega(\cdot)=\exp(\textbf{i}\cdot)$. We let $\mu$ be a measure with a symmetric Fourier transform. So, the estimator of $\phi_0$ will be
\begin{equation}\label{eq: np estimator with beta=0}
\widehat{\phi}=a  \sum_{l=0}^m (I-a \widehat{A}^*_Z \widehat{A}_Z)^l \widehat{A}^*_Z \widehat{s}\, ,
\end{equation}
where $\widehat A_Z$ is defined in (\ref{eq: def of A hat_Z}) and $\widehat s$ in (\ref{eq: def of s hat}). To obtain a more explicit expression for $\widehat{\phi}$, let us compute the compositions $\widehat{A}_Z^* \widehat{A}_Z$ and $\widehat{A}_Z^* \widehat{s}$. By the expressions of $\widehat{A}_Z$, $\widehat{A}_Z^*$, and $\widehat{s}\,$  in previous section, we get (see the comments below) 
\begin{align}\label{eq: AhatstarZ shat}
    (\widehat{A}_Z^* \widehat{s})(z)=& \int_{ }\frac{1}{n} \sum_{i=1}^n  Y_i\,  \omega(W_i^T t) \,  \overline{\widehat{k}}(z,t)\, \frac{1}{\pi(z)}\,  \mu(d t) \nonumber \\
    =& \frac{1}{n^2 h^p}\sum_{i,j=1}^n Y_i \int_{ }\, \omega(W_i^T t)\, \overline{\omega}(W_j^T t) \, \mu(d t) \, K\left( \frac{z-Z_j}{h} \right)\, \frac{1}{\pi(z)} \nonumber \\
    =& \frac{1}{n^2 h^p} \sum_{i,j=1}^n Y_i \,\int_{ }\, \exp(\textbf{i}(W_i-W_j)^T t)\,  \mu(d t )\, K\left( \frac{z-Z_j}{h}\right)\, \frac{1}{\pi(z)} \nonumber \\
    =& \frac{1}{n^2 h^p} \sum_{i,j=1}^n Y_i \,\mathcal{F}_{\mu}(W_i-W_j)\, K\left( \frac{z-Z_j}{h}\right) \, \frac{1}{\pi(z)} \, ,
\end{align}
 where in the second equality we have used the expression of $\widehat{k}$ from (\ref{eq: def of k hat}), while in the fourth equality $\mathcal{F}_\mu$ denotes the characteristic function of the finite measure $\mu$. Pick any $\varphi\in L^2_\pi(\mathbb{R}^p)$. By steps similar to those in the previous display, we find
\begin{align}\label{eq: AhatstarZ AhatZ varphi}
    (\widehat{A}_Z^* \widehat{A}_Z \varphi)(z)=& \int_{ }\,(\widehat{A}_Z \varphi)(t)\, \overline{\widehat{k}}(z,t)\,\frac{1}{\pi(z)}\, \mu(d t )\nonumber \\
    =& \frac{1}{n h^{p}} \,\sum_{i=1}^n \,K\left(\frac{z-Z_i}{h}\right)\, \frac{1}{\pi(z)}\,  \int_{ }\,(\widehat{A}_Z \varphi)(t)\, \overline{\omega}(W_i^T t)\, \mu(d t ) \nonumber \\
    =& \frac{1}{n^2 h^{2 p}} \sum_{i,j=1}^n K\left(\frac{z-Z_i}{h}\right)\, \frac{1}{\pi(z)}\,\nonumber \\
    & \cdot \int_{ }\, \omega(W_j^T t) \, \overline{\omega}(W_i^T t) \, \mu(d t )\, \int_{ }\, K\left(\frac{z_2-Z_j}{h}\right)\, \varphi(z_2) d z_2  \nonumber \\
    =& \frac{1}{n^2 h^{2 p}} \sum_{i,j=1}^n K\left(\frac{z-Z_i}{h}\right)\, \frac{1}{\pi(z)}\,\nonumber  \\
    &\cdot \,  \mathcal{F}_\mu(W_i-W_j) \int_{ }\,  \,K\left(\frac{z_2-Z_j}{h}\right) \varphi(z_2) d z_2 \, ,
\end{align}
where in the last equality we have used the symmetry of $\mathcal{F}_\mu$. 
We will now compare the above expressions with the approach that smooths on the IVs, see, e.g., \cite{carrasco_chapter_2007} and \cite{florens_instrumental_2012}. To briefly summarize such an approach, we start from $\mathbb{E}\{U|W\}=0$ (where $\beta_0\equiv 0$). We multiply both sides to this equation by $f_W$ (the density of $W$) and obtain the integral equation  
\begin{equation}\label{eq: integral equation for the classical approach}
    r=T_Z \phi_0\, ,
\end{equation}
where
\begin{equation*}
r(\cdot):=\mathbb{E}\{Y|W=\cdot\}f_W(\cdot) \, ,
\end{equation*}
\begin{equation}\label{eq: def of T_Z}
 T_Z \varphi:=\int \varphi(z) f_{W Z}(\cdot,z) d z \, ,\,T_Z:L^2_\pi(\mathbb{R}^p)\mapsto L^2(\mathbb{R}^q) 
\end{equation}
 and $f_{W Z}$ denotes the joint density of $(W,Z)$. The Hilbert adjoint of $T_Z$ is  
 \begin{equation*}
     T_Z^* \psi:=\int \psi(w) f_{W Z}(w,\cdot) dw \, /\pi(\cdot) \, ,\,T_Z^*:L^2(\mathbb{R}^q)\mapsto L^2_\pi (\mathbb{R}^p)\, .
 \end{equation*}
 To estimate $r$, $T_Z$, and $T_Z^*$, we let $h_W$ be a bandwidth and $K_W$ be a kernel. Then, the joint density  $f_{W Z}$ is estimated as 
 \begin{equation*}
     \widehat{f}_{W Z}(w,z):=\frac{1}{n h^p h_W^q} \sum_{i=1}^n K\left(\frac{z-Z_i}{h}\right)\, K_W\left(\frac{w-W_i}{h}\right)
 \end{equation*}
 and $r,T_Z,T^*_Z$ are estimated as 
 \begin{align}\label{eq: definition of the estimators in the classical approach}
     \widehat{r}(w):= & \frac{1}{n h^q_W} \sum_{i=1}^n Y_i K_W\left(\frac{w-W_i}{h}\right)\, , \nonumber  \\
     (\widehat{T}_Z\varphi)(w):=&\int_{ }\varphi(z)\widehat{f}_{W,Z}(w,z)d z\, , \nonumber  \\
     \text{ and }\, (\widehat{T}^*_Z\psi)(z):&=\int_{ }\psi(w) \widehat{f}_{W,Z}(w,z) \frac{1}{\pi(z)} d w\, . 
 \end{align}
 We can define the estimator from the approach that smooths on the IVs as 
 \begin{equation}\label{eq: np estimator of the classical approach}
     \widetilde{\phi}:=a\sum_{l=0}^m (I-a\widehat{T}^*_Z \widehat{T}_Z)^l\widehat{T}^*_Z \widehat{r}\, .
 \end{equation}
Now, given the above estimators, we can compute the compositions $\widehat{T}^*_Z\widehat{r}$ and $\widehat{T}^*_Z \widehat{T}_Z$. We have 
\begin{align}\label{eq: ThatstarZ rhat}
    (\widehat{T}^*_Z \widehat{r})(z)=& \int_{ }\frac{1}{n h_W^q} \sum_{i=1}^n Y_i K_W\left(\frac{w-W_i}{h_W}\right)\, \widehat{f}_{W Z}(w, z) \frac{1}{\pi(z)} d w \nonumber \\
    =& \frac{1}{n^2 h^p  } \sum_{i,j=1}^n Y_i h_W^{-2 q}\int_{ }K_W\left(\frac{w-W_i}{h_W}\right)\, K_W\left(\frac{w-W_j}{h_W}\right)\, d w\, \nonumber 
    \cdot K\left(\frac{z-Z_j}{h}\right)\, \frac{1}{\pi(z)} \nonumber \\
    =& \frac{1}{n^2 h^p } \sum_{i,j=1}^n Y_i \frac{1}{h_W^q}\, (K_W \ast K_W)\left(\frac{W_j-W_i}{ h_W}\right)\, K\left(\frac{z-Z_j}{h}\right)\, \frac{1}{\pi(z)}  \, ,
\end{align}
where in the last equality $K_W\ast K_W$ denotes the convolution of $K_W$ with itself and we have used a classical change of variable. Also, for any $\varphi\in L^2_\pi(\mathbb{R}^p)$ we have 
\begin{align}\label{eq: ThatstarZ ThatZ varphi}
    (\widehat{T}^*_Z & \widehat{T}_Z\varphi)(z)=\int_{ }\,(\widehat{T}_Z\varphi)(w)\, \widehat{f}_{W Z}(w,z)\, d w \frac{1}{\pi(z)} \nonumber \\
     =& \frac{1}{n h^p h_W^q} \sum_{i=1}^n K\left(\frac{z-Z_i}{h}\right)\frac{1}{\pi(z)}\,  \int_{ } \, (\widehat{T}_Z\varphi)(w)\, K_W\left(\frac{w-W_i}{h_W}\right)\, d w \nonumber \\
    =& \frac{1}{n^2 h^{2 p} } \sum_{i,j=1}^n K\left(\frac{z-Z_i}{h}\right)\frac{1}{\pi(z)} \nonumber \\
    &\cdot h_W^{-2 q} \int_{ } K_W\left(\frac{w-W_i}{h_W}\right)\, K_W\left(\frac{w-W_j}{h_W}\right)\, d w \int_{ }K\left(\frac{z_2-Z_j}{h}\right)\, \varphi(z_2) \, d z_2 \nonumber  \\
    =& \frac{1}{n^2 h^{2 p}} \sum_{i,j=1}^n K\left(\frac{z-Z_i}{h}\right)\frac{1}{\pi(z)} \,\nonumber \\
    &\cdot \frac{1}{h_W^q}\,  (K_W \ast K_W)\left(\frac{W_j-W_i}{ h_W}\right)\,\int_{ }K\left(\frac{z_2-Z_j}{h}\right)\, \varphi(z_2)\, d z_2\, , 
\end{align}
 where in the last equality we have used a classical change of variable.\\
Let us now compare (\ref{eq: AhatstarZ shat}) with (\ref{eq: ThatstarZ rhat}) and (\ref{eq: AhatstarZ AhatZ varphi}) with (\ref{eq: ThatstarZ ThatZ varphi}). We realize that as long as 
\begin{equation}\label{eq: convolution and fourier transform}
\frac{1}{h_W^q}\, (K_W\ast K_W)\left(\frac{\cdot}{h_W}\right)=\mathcal{F}_\mu    \, ,
\end{equation}
we have  $\widehat{A}^*_Z\widehat{s}=\widehat{T}^*_Z\widehat{r}$ and $\widehat{A}^*_Z\widehat{A}_Z=\widehat{T}^*_Z \widehat{T}_Z$. By comparing Equations (\ref{eq: np estimator with beta=0}) and  (\ref{eq: np estimator of the classical approach}) we obtain that $\widehat{\phi}=\widetilde{\phi}$. So if the bandwidth $h_W$ is kept fixed and the above equality is satisfied, the estimator from the classical approach that smooths on the IVs will equal our proposed estimator.
\footnote{ When $\widehat{A}^*_Z \widehat{s}=\widehat{T}_Z \widehat{r}$ and $\widehat{A}^*_Z \widehat{A}_Z=\widehat{T}^*_Z \widehat{T}_Z$,
the correspondence between our approach and the one that smooths over the IVs will remain valid also with a Tikhonov regularization scheme. This is because the Tikhonov regularization depends only on the compositions $\widehat{A}^*_Z \widehat{A}_Z$ and $\widehat{A}^*_Z \widehat{s}$. } \\
Equation (\ref{eq: convolution and fourier transform}) can be ensured in several cases. As an example, assume that $K_W$ is a product kernel between $q$ standard Gaussian densities. Then $(K_W \ast K_W)(\cdot)=2^{-q/2}K_W(\cdot/\sqrt{2} )$. So, if $\mu$ is set equal to the product between $q$ standard Gaussian densities each divided by $\sqrt{2 \pi}$, then the equality in (\ref{eq: convolution and fourier transform}) will hold as long as $h_W=1/\sqrt{2}$. \\
We finally remark several differences between our approach that does not smooth on the IVs and the classical approach that smooths on the IVs. First, the approach we propose treats in the same way the case where $W$ and $Z$ have common components and the case where they don't. Specifically, in either of such cases, from Equations (\ref{eq: def of A_Z}) and (\ref{eq: AZ as an integral operator}) $A_Z$ remains an  integral/kernel operator, so it remains bounded and continuous.\footnote{From Footnote \ref{footnote: about the definition of kernel operator, compactness, and non continuous inverse}, $A_Z$ is a compact operator both when $Z$ and $W$ have common components and when they don't. From  \cite[Theorems 2.5 and 2.14]{kress_linear_2012}, compact operators are bounded and continuous.}\footnote{Estimating a kernel operator is statistically convenient, as it boils down to estimating its kernel. In particular, estimating $A_Z$ is equivalent to estimating $k$, see Equation (\ref{eq: AZ as an integral operator}).}  
This is a feature not shared by the approach that smooths on the IVs. Specifically, when $W$ and $Z$ have common components, from (\ref{eq: def of T_Z})  $T_Z$ will no longer be a kernel operator, see Footnote \ref{footnote: about the definition of kernel operator, compactness, and non continuous inverse}, and will not be neither continuous nor compact, see \cite{carrasco_chapter_2007}.
To overcome this problem, the usual approach is to fix the values of the common components between $Z$ and $W$ and to conduct the estimation locally to such values. This guarantees that {\itshape locally to such values} $T_Z$ is a kernel operator and hence compact and continuous, see, e.g. \cite{darolles_nonparametric_2011} or \cite{hall_nonparametric_2005}. Differently, from (\ref{eq: def of A_Z}) the operator $A_Z$ remains a kernel operator both in the case where $W$ and $Z$ share common components and in the case where they don't. So, within our approach, if such common components are present, we do not need to fix them and conduct the estimation locally to such values.\\
Second, the estimation of $A_Z$ does not require smoothing over the instruments $W$, so we will not need to select a smoothing parameter  for the IVs. This is because estimating $A_Z$ boils down to computing $\widehat k$ in (\ref{eq: def of k hat})  that only requires a smoothing parameter for $Z$. Differently, the classical approach requires selecting a smoothing parameter for the IVs {\itshape and } for $Z$, see (\ref{eq: definition of the estimators in the classical approach}).\\
Third, the left hand side of the integral equation (\ref{eq: Bierens equation in terms of AX and AZ}) in our approach, i.e. $s$, is estimated as an empirical average and hence at a parametric rate, see (\ref{eq: def of s hat}). Differently, for the approach that smooths on the IVs, the left hand side of the integral equation in (\ref{eq: integral equation for the classical approach}), i.e. $r$, is nonparametrically estimated and requires selecting an additional bandwidth.

\section{Assumptions and Asymptotic Behavior}\label{sec: Assumptions and Asymptotic Behavior}
For presentation purposes, we first study the asymptotics of our estimator in the fully nonparametric model, i.e. when $\beta_0\equiv 0$. Then, we obtain the asymptotics for the estimators of the semiparametric partly linear model.

\subsection{The Fully Nonparametric Model}
Let us consider the fully nonparametric model
\begin{equation}\label{eq: fully np model}
    Y = \phi_0(Z)+U \text{ with }\mathbb{E}\{U|W\}=0\, .
\end{equation}
This is just a specific case of the more general model (\ref{eq: PL NPIV model}) with $\beta_0\equiv 0$. 
Then, by (\ref{eq: Bierens equation}) we have 
\begin{equation}\label{eq: integral equation for the fully np model}
    s=A_Z \phi_0\, .
\end{equation}
The estimator $\widehat \phi$ will be as in Equation (\ref{eq: np estimator with beta=0}).
In this section, we first state the assumptions. Then, we obtain the convergence rate for $\widehat{\phi}$ and the asymptotic normality of the inner product involving $\widehat \phi$.\\

Let $\mathcal{N}(A_Z)$ denote the null space of $A_Z$ and let $\mathcal{N}(A_Z)^\perp$ be its orthogonal complement.\footnote{Formally, $\mathcal{N}(A_Z):=\{\varphi\in L^2_\pi(\mathbb{R}^p)\,:\,A_Z \varphi=0\}$ and $\mathcal{N}(A_Z)^\perp:=\{\phi\in L^2_\pi(\mathbb{R}^p)\,:\,\left<\phi,\varphi\right>=0\text{ for all }\varphi\in \mathcal{N}(A_Z)\}$. } We denote with $\phi_0^\perp$ the projection of $\phi_0$ onto $\mathcal{N}(A_Z)^\perp$ and let $U^\perp:=Y-X^T\beta_0-\phi_0^\perp(Z)$. 

\begin{hp}\label{ass: iid and omega}
(i) $(Y_i,X_i,Z_i,W_i, U_i, U_i^\perp)_{i=1}^n$ is an iid sample, $Y,\, U,U^\perp$, and each component of $X$ have finite second moments, the support of $W$ is bounded; (ii) $\omega:\mathbb{R}\mapsto \mathbb{C}$ is an analytic non-polynomial function with $\partial^l \omega (0)\neq 0$ for each $\l\in \mathbb{N}$ and $\sup_{t\in\mathcal{T}}|\omega(W^T t)|\leq C$ for a constant $C$;  (iii) $\mathcal{T}\subset \mathbb{R}^q$ is a set containing a neighborhood of the origin; (iv) the space $L^2_\mu(\mathcal{T})$ is separable and $\mu$ is a positive finite measure on $\mathcal{T}$.
\end{hp}

We now define the following class of functions which is needed to state the integrability and smoothness conditions, see \cite{delgado_significance_2001} and \cite{florens_instrumental_2012}.

\begin{definition}
For a given function $\gamma$ and for $\alpha \geq 0$, $\upsilon>0$, the space $\mathcal{B}^{\upsilon, \alpha}_\gamma (\mathbb{R}^\ell)$ is the class of functions $g:\mathbb{R}^\ell \mapsto \mathbb{R} $ satisfying: (i) $g$ is everywhere $(b-1)$ times differentiable for $b-1 < \upsilon \leq b $; (ii) for some $R>0$ and for all $x$, the inequality 
$$\sup_{y : \|y-x\|<R}\frac{\left|g(y)-g(x)-Q(y-x)\right|}{\|y-x\|^\upsilon}\leq \eta(x)$$
holds true, where $Q=0$ when $b=1$, while when $b>1$ $Q$ is a $(b-1)$ degree homogeneous polynomial in $(y-x)$ with coefficients the partial derivatives of $g$ at $x$ of orders $1$ through $b-1$; $\eta$ is a function uniformly bounded by a constant when $\alpha=0$, while when $\alpha>0$ the functions $g$ and $\eta$ are such that $g^\alpha / \gamma\, , \eta^\alpha / \gamma  \in L^1(\mathbb{R}^\ell)$.
\end{definition}

Let us denote with $f_{W Z}$ the joint density of $W$ and $Z$ with respect to the Lebesgue measure.

\begin{hp}\label{ass: smoothness and integrability}
 $f_Z \in \mathcal{B}^{1,2}_\pi (\mathbb{R}^p)\cap \mathcal{B}^{1,1}_\pi(\mathbb{R}^p)$; $\phi_0, \phi_0^\perp \in \mathcal{B}^{\rho,0}_\pi(\mathbb{R}^p)\cap L^2_\pi(\mathbb{R}^p)$; $f_{W Z}\in\mathcal{B}_{\pi\,1}^{\rho,2}(\mathbb{R}^p\times \mathbb{R}^q)$. 
\end{hp}

\begin{hp}\label{ass: Kernel K}
The kernel $K$ is symmetric about $0$ and of order $\rho$, with $\int_{}K(u)^2 |u|<\infty$.  $K((\cdot-Z_i)/h)/\pi\in L^2_\pi (\mathbb{R}^p)$ for all $i=1,\ldots,n$.
\end{hp}

We now introduce the source condition on $\phi_0$. To this end, let $(\lambda_j, \varphi_j,\psi_j)_j$ be the singular system of $A_Z$, where $(\lambda_j)_j$ is a sequence of values in $\mathbb{R}_{++}$, $(\varphi_j)_j$ is a sequence of orthonormal elements in $L^2_\pi(\mathbb{R}^p)$, and $(\psi_j)_j$ is a sequence of orthonormal elements in $L^2_\mu(\mathcal{T})$. $(\lambda_j, \varphi_j,\psi_j)_j$ satisfy 
$$A_Z\varphi_j=\lambda_j \psi_j\,\text{ and }\,A_Z^*\psi_j=\lambda \varphi_j\, ,$$
see \cite[Definition 15.15 and Theorem 15.16]{kress_linear_2012}.
\begin{hp}\label{ass: source condition}
 For some $\iota > 0$ : $ \sum_j \lambda_j^{-2 \iota} \left|\left< \phi_0, \varphi_j \right> \right|^2  <\infty\, . $
\end{hp}
Assumption \ref{ass: iid and omega} formally states the conditions on $\omega$ and $\mathcal{T}$ which guarantee that the continuum of moment conditions in (\ref{eq: continuum of unconditional moments}) is equivalent to $\mathbb{E}\{U|W\}=0$. As already detailed in the previous sections, different choices are available for $\omega$. The boundedness of  $\omega$ is needed to obtain the convergence rates of $\widehat{s}$ and $\widehat{A}_X$ using CLTs for Hilbert-valued random elements. The square integrability conditions in Assumption \ref{ass: iid and omega} are standard in the literature. \footnote{The square integrability of $U^\perp$ will be used when obtaining the asymptotic normality of $\sqrt{n}(\widehat \beta-\beta_0)$ without the injectivity of $A_Z$.}
Assumption \ref{ass: smoothness and integrability} imposes the smoothness and integrability conditions on the nonparametric functions $\phi_0$ and $\phi_0^\perp$, on the density $f_Z$, and on the joint density $f_{W Z}$. These smoothness conditions have to be linked to the order of the kernel $\rho$ in Assumption \ref{ass: Kernel K} to control the bias of the nonparametric estimator $\widehat{A}_Z$.\footnote{The smoothness conditions on $\phi_0^\perp$ will be used when obtaining the asymptotic normality of $\sqrt{n}(\widehat{\beta}-\beta_0)$ without the injectivity of $A_Z$.}
Assumption \ref{ass: source condition} is common in the inverse problems literature, see, e.g.  \cite{darolles_nonparametric_2011}, \cite{carrasco_chapter_2007}, \cite{hall_nonparametric_2005}, and \cite{engl2000regularization}. 
It contains a source condition on $\phi_0$. $\iota$ measures the degree of ill-posedness, in the sense that the smaller the $\iota$ the more the inverse problem in (\ref{eq: integral equation for the fully np model}) will be ill-posed. 
 $\iota$ can be interpreted as the degree of smoothness of $\phi_0$, see \cite{carrasco_chapter_2007}.

\begin{thm}\label{prop: convergece rate of phihat in the np model}
Consider the fully nonparametric model in (\ref{eq: fully np model}) and the associated integral equation (\ref{eq: integral equation for the fully np model}). Let the estimator of the fully nonparametric model be 
$$\widehat{\phi}= a  \sum_{l=0}^m (I-a \widehat{A}^*_Z \widehat{A}_Z)^l \widehat{A}^*_Z \widehat{s}\, .$$
Also, define the regularized solution to the integral equation (\ref{eq: integral equation for the fully np model}) as 
$$\phi_m=a \sum_{l=0}^m (I-a A_Z^* A_Z)^l A_Z^* s\, .$$
Then, under Assumptions \ref{ass: identification}\textcolor{blue}{(b)}, \ref{ass: iid and omega}, \ref{ass: smoothness and integrability}, and \ref{ass: Kernel K} we have 
\begin{equation}\label{eq: rate for the np model without source condition}
    \|\widehat{\phi}-\phi_0\| =O_P\left( m\,b_n\,+ \left[1+ m a_n\right]\, \|\phi_m - \phi_0\|\,\right)\,,
\end{equation}
where $b_n=n^{-1/2}+h^\rho$ and $a_n=(n h^p)^{-1/2}+h^\rho$ . 
If moreover Assumption \ref{ass: source condition} holds then $\|\phi_m-\phi_0\|\lesssim m^{-\iota/2}$ and\footnote{Given two sequences $(a_m)_m$ and $(b_m)_m$, $a_m\lesssim b_m$ means that $a_m\leq C b_m$ for a universal constant $C$.}
\begin{equation}\label{eq: rate for the np model with source condition}
    \|\widehat \phi - \phi_0\|=O_P\left(\, m\,b_n\,+ \left[1+ m a_n\right]\, m^{-\iota/2}\,\right) .
\end{equation}
\end{thm}
From \citet[Definition 15.5 and Theorem 15.27]{kress_linear_2012}, under Assumptions \ref{ass: identification}\textcolor{blue}{(b)} and \ref{ass: iid and omega} we have  $\|\phi_m-\phi_0\|=o(1)$ as $m\rightarrow \infty$.
Thus, the previous theorem directly implies the consistency of $\widehat \phi$.

\begin{cor}\label{corollary: consistency of phihat}
    Under Assumptions \ref{ass: identification}\textcolor{red}{(b)}, \ref{ass: iid and omega}, \ref{ass: smoothness and integrability}, and \ref{ass: Kernel K},  if (i) $n h^p/m^2\rightarrow \infty$ and (ii) $m h^\rho=o(1)$, then 
    \begin{equation*}
        \|\widehat \phi - \phi_0\|=o_P(1)\, .
    \end{equation*}
\end{cor}
It is useful to briefly discuss how $p$ (the dimension of the regressors $Z$)  and $\rho$ (the kernel order) must be related to guarantee the conditions of the previous corollary and hence the consistency of $\widehat \phi$. If $m\sim n^\alpha$ and $h\sim n^{-\gamma}$ for $\gamma,\alpha>0$, then the condition $n h^p/m^2\rightarrow \infty$ is equivalent to $0<\gamma<(1-2\alpha)/p$ and  $m h^\rho=o(1)$ is equivalent to $\rho>\alpha/\gamma$. Thus, for a given $\alpha\in(0,1/2)$, i.e. for a given convergence rate of the regularization parameter $m$, the larger the dimension $p$ of $Z$ the larger must the kernel order $\rho$ to guarantee that $\widehat \phi$ is a consistent estimator for $\phi_0$. \\
Beyond showing the consistency of our estimator, Corollary \ref{corollary: consistency of phihat} allows us to discover a very interesting property of $\widetilde \phi$ in (\ref{eq: np estimator of the classical approach}), i.e. the estimator of the classical approach that smooths on the IVs.  As we have shown in Section \ref{sec: A fixed bandwidth interpretation}, our estimator $\widehat \phi$ coincides with the classical estimator $\widetilde \phi$ in (\ref{eq: np estimator of the classical approach}) that smooths over the IVs but keeps the bandwidth on the instruments ($h_W$) fixed as the sample size increases. Clearly, as long as the bandwidth $h_W$ remains fixed as the sample size increases, the nonparametric estimators $\widehat T_Z$, $\widehat T_Z^*$, and $\widehat r$ in (\ref{eq: definition of the estimators in the classical approach}) will {\itshape not} be consistent, as their nonparametric bias will {\itshape not} vanish. However, since for $h_W$ fixed  $\widetilde \phi$ coincides with $\widehat \phi$, Corollary \ref{corollary: consistency of phihat} {\itshape shows that $\widetilde{\phi}$ will remain consistent for $\phi_0$ although it will be based on the estimators $\widehat{r}$, $\widehat{T}_Z$, and $\widehat{T}_Z^*$ that are not consistent for their population counterparts.} \\
These findings have a similar flavour to those  in \cite{escanciano2018simple} and \cite{lavergne_smooth_2013} obtained in a parametric context. When a {\itshape finite dimensional} parameter is identified by a conditional moment restriction, \cite{escanciano2018simple} and \cite{lavergne_smooth_2013} show that a parametric estimator minimizing the distance between the estimated conditional moment and zero remains consistent although the bandwidth used to estimate such a conditional moment is kept fixed with the sample size. Our finding has a similar flavor, but our context is substantially different: here we are proving such a result for an {\itshape infinite} dimensional estimator ($\widehat \phi$) and within an {\itshape ill-posed inverse problem}.\\
Finally, Equation (\ref{eq: rate for the np model with source condition}) in Theorem \ref{prop: convergece rate of phihat in the np model} shows the convergence rate of $\widehat \phi$ under the source condition in Assumption \ref{ass: source condition}. Clearly, the faster the regularization bias $\|\phi_m-\phi_0\|$ goes to zero, i.e. the larger $\iota$, the faster $\widehat \phi$ can converge towards $\phi_0$. We remark that thanks to the Landweber-Fridman scheme the convergence rate of $\|\widehat \phi - \phi_0\|$ can exploit the full smoothness of $\phi_0$ represented by $\iota$. Differently, if the Tikhonov regularization was used a degree of smoothness larger than or equal to 2 would not make any difference on the convergence rate of $\widehat \phi$, see \cite[Propositions 3.11 and 4.1]{carrasco_chapter_2007}.\\  
We do not claim that the convergence rate provided in (\ref{eq: rate for the np model with source condition}) is sharp. An in-depth discussion about optimality of the convergence rates in general ill-posed inverse problems is provided in \cite[Proposition 4.2 and pages 5686-5687]{carrasco_chapter_2007}.\footnote{Specifically, the optimal convergence rate of $\widehat \phi$ can be obtained from \cite[Proposition 4.2]{carrasco_chapter_2007}}  \\


We complete the analysis of the fully nonparametric model by studying the $\sqrt{n}$ asymptotic normality of the inner product involving $\widehat \phi$. 

\begin{thm}\label{prop: asynorm of the innder product}
Let Assumptions \ref{ass: identification}\textcolor{red}{(b)} ,\ref{ass: iid and omega}, \ref{ass: smoothness and integrability} ,  \ref{ass: Kernel K},   
hold. Let $m$ and $h$ be such that $n/m^3=o(1)$, $n h^p/m^2\rightarrow \infty$, and $n h^{2 \rho}=o(1)$. Also, let $\phi_0$ satisfy Assumption \ref{ass: source condition} with $\iota\geq 2$ and
let $g\in L^2_\pi(\mathbb{R}^p)$ be such that 
\begin{equation*}
    \sum_j \frac{\left|\left<g,\varphi_j\right>\right|^2}{\lambda_j^{2 \gamma}}<\infty\, 
\end{equation*}
with $\gamma \geq 2$. Then,
\begin{equation*}
    \sqrt{n}\,\left<\widehat \phi - \phi_0, g\right>=n^{-1/2}\sum_{i=1}^n U_i\left<A_Z^*[\omega(W_i^T \cdot)] , (A_Z^* A_Z)^{-1} g \right>+o_P(1)\, .
\end{equation*}
\end{thm}
The conditions on $m$ and $h$ in Theorem \ref{prop: asynorm of the innder product}  are stronger than those required for the consistency of $\widehat \phi$ in Corollary \ref{corollary: consistency of phihat}.\footnote{To see that the conditions on $m$ and $h$ in Theorem \ref{prop: asynorm of the innder product} are stronger than those in Corollary \ref{corollary: consistency of phihat}, notice that for $h\rightarrow 0$ the conditions $n h^p/m^2\rightarrow \infty$ and  $n h^{2\rho}=o(1)$  imply $m h^\rho=\sqrt{m^2/n}\sqrt{n h^{2\rho}}=o(1)$.} To clarify briefly how the kernel order $\rho$, the dimension $p$, and the regularization parameter $m$ must be linked to satisfy the conditions of Theorem \ref{prop: asynorm of the innder product}, let us assume that $m\sim n^\alpha$ and $h\sim n^{-\gamma}$ for $\alpha,\gamma>0$. Then, the conditions in Theorem \ref{prop: asynorm of the innder product} will be satisfied for $1/3<\alpha<1/2$,  $\gamma<(1-2\alpha)/p$, and $\rho>1/(2\gamma)$. Thus, for a given $\alpha$ (i.e. for a given convergence rate of the regularization parameter $m$), the larger the dimension $p$ of the regressors $Z$ the larger must be the kernel order $\rho$ to satisfy the conditions of Theorem \ref{prop: asynorm of the innder product}.
The condition $\sum_j \left|\left<g,\varphi_j\right>\right|^2\lambda_j^{-2 \gamma}<\infty$ is a source condition on $g$ and is similar in nature to the source condition imposed on $\phi_0$ in Assumption \ref{ass: source condition}. Intuitively, it connects the ``smoothness" of $g$ (measured by the rate of decay of the Fourier coefficients $\left<g,\varphi_j\right>$) with the degree of ill-posedness of the inverse problem (measured by the decay of the singular values $\lambda_j$), see \cite{darolles_nonparametric_2011}. Such a condition ensures that $\|(A_Z^*A_Z)^{-1} g\|<\infty$. Without this condition, the asymptotic variance of $\sqrt{n}\,\left<\widehat \phi - \phi_0, g\right>$ would be  infinite. As a consequence, the convergence rate of $\left<\widehat \phi - \phi_0, g\right>$ would be slower than $\sqrt{n}$, and we would have to normalize $\left<\widehat \phi - \phi_0, g\right>$ by a different rate to obtain the asymptotic normality. For a discussion of this case, we refer the reader to \cite[Chapter 3]{racine2014oxford}.\footnote{The study of the convergence in distribution of $\left<\widehat \phi - \phi_0, g\right>$ when $\|(A_Z^* A_Z)^{-1} g\|=\infty$ is beyond the scope of this paper. }  

\subsection{The Semiparametric Model}\label{sec: The semiparametric model}
In this section we study the estimation of the semiparametric partly linear model 
\begin{equation*}
    Y=X^T\beta_0 + \phi(Z)+U \text{ with }\mathbb{E}\{U|W\}=0\, .
\end{equation*}
As seen in Section \ref{sec: The proposed approach and identification}, this model gives rise to the integral equation
\begin{equation*}
    s=A_X \beta_0 + A_Z \phi_0\, ,
\end{equation*}
where $s$, $A_X$, and $A_Z$ are defined in (\ref{eq: definition of s}), (\ref{eq: definition of AX}), and (\ref{eq: def of A_Z}). Our goal in this section is to obtain the $\sqrt{n}$ asymptotic normality of $\widehat \beta$ in (\ref{eq: definition of betahat}) and the convergence rates for $\widehat \phi$ in (\ref{eq: definition of phihat}) based on the Landweber-Fridman regularization scheme. 
 Let $(\mu_j, e_j, \widetilde{\psi}_j)_{j=1}^\kappa$ be the singular system of $A_X$, where  $(\mu_j)_{j=1}^\kappa$ is a collection of values in $\mathbb{R}_{++}$, $(e_j)_{j=1}^\kappa$ is a collection of orthonormal elements in $\mathbb{R}^{\kappa}$, and $(\widetilde{\psi}_j)_{j=1}^\kappa$ is a collection of orthonormal elements in $L^2_\mu(\mathcal{T})$. $(\mu_j, e_j, \widetilde{\psi}_j)_{j=1}^\kappa$ satisfy 
 $$A_X e_j=\mu_j \widetilde{\psi}_j\,\text{ and }A_X^* \widetilde{\psi}_j=\mu_j e_j\, ,$$
 see \cite[Definition 15.15 and Theorem 15.16]{kress_linear_2012} . We introduce the following source condition on the eigenvectors of $A_X$:\footnote{Indeed, since $A_X$ is defined over $\mathbb{R}^\kappa$, its range will have dimension at most equal to $\kappa$. Thus, its singular system will have at most $\kappa$ elements,  see \cite{kress_linear_2012}  . }

 \begin{hp}\label{ass: source condition on A x}
  For $\eta\geq 2$ : $\max_{\ell=1,\ldots,\kappa}\sum_{j} \lambda_j^{-2 \eta}\left|\left< \widetilde{\psi}_\ell , \psi_j \right>\right|^2<\infty\, .$
 \end{hp}
The above source condition on the eigenvectors of $A_X$ is similar in spirit to Assumption 3.1 in \cite{florens_instrumental_2012}. The parameter $\eta$ can be interpreted as a measure of the degree of ``orthogonality" between $A_X$ and $A_Z$. In fact, when the ranges of $A_Z$ and $A_X$ are orthogonal, $\left<\psi_j,\widetilde{\psi}_\ell\right>=0$ for all $j,\ell$, and hence $\eta=\infty$. Thus, the larger the parameter $\eta$ the more ``orthogonal" are the ranges of $A_Z$ and $A_X$. For a more in depth discussion of Assumption \ref{ass: source condition on A x} see \cite{florens_instrumental_2012}.  
When it comes to the estimation of $\widehat \beta$, Assumption \ref{ass: source condition on A x} and Assumption \ref{ass: source condition}
ensure that the estimation of the projection operator $P_Z$ will not impact the influence-function representation of $\widehat{\beta}-\beta_0$.

Theorem \ref{prop: asy norm of betahat and convergence rates of phihat in the PL model} below obtains the asymptotic normality of $\sqrt{n}(\widehat{\beta}-\beta_0)$ and the convergence rate of $\widehat{\phi}$ in the semiparametric partly-linear model. We let 
\begin{equation*}
    \widehat{\phi}= a  \sum_{l=0}^m (I-a \widehat{A}^*_Z \widehat{A}_Z)^l \widehat{A}^*_Z (\widehat{s}-\widehat{A}_X \widehat{\beta})
\end{equation*}
and we define
\begin{align*}
    &\Psi :=  \Sigma^{-1} \mathbb{E}\left\{ \sigma(W)^2 \,G(W)\,G(W)^T \right\} \Sigma^{-1} \, ,\\
   &\sigma(W)^2:=\mathbb{E}\{(U^\perp)^2|W\}\, ,    \\
    &G(w):=\mathbb{E}\left\{X \int_{ }\overline{\omega}(W^T t) \left[(I-P_Z)\omega(w^T \cdot) \right](t)\, \mu(d t)\right\} \, .
\end{align*}
\begin{thm}\label{prop: asy norm of betahat and convergence rates of phihat in the PL model}
Let Assumptions \ref{ass: identification}\textcolor{red}{(a)}, \ref{ass: iid and omega}, \ref{ass: smoothness and integrability}, \ref{ass: Kernel K}
hold. \\

\begin{enumerate}
        \item[(i)] Let $\phi_0$ satisfy Assumption \ref{ass: source condition} with $\iota\geq 2$ and let Assumption \ref{ass: source condition on A x} hold. Also, let $m$ and $h$ satisfy $n/m^3=o(1)$, $n h^p/m^2\rightarrow \infty$, and $n h^{2\rho}=o(1)$. Then,  
\begin{equation}\label{eq: IFR for betahat}
    \sqrt{n}(\widehat{\beta}-\beta_0)=\frac{1}{\sqrt{n}}\sum_{i=1}^n U_i^\perp \Sigma^{-1} A_X^* (I-P_Z)\omega(W_i^T\cdot) + o_P(1) \leadsto \mathcal{N}(0,\Psi)\, .
\end{equation}
If moreover Assumption \ref{ass: identification}\textcolor{blue}{(b)} holds, 
we have 
\begin{equation}\label{eq: rate for phihat in the PL model}
    \|\widehat{\phi}-\phi_0\|=O_P\left(   m\,b_n\,+ \left[1+ m a_n\right]\, m^{-\iota/2}\,\right)\, ,
\end{equation}
where $b_n=n^{-1/2}+h^\rho$ and $a_n=(n h^p)^{-1/2}+h^\rho$ .\\

\vspace{0.25cm}

\item[(ii)] Let $\phi_0$ satisfy Assumption \ref{ass: source condition} with $\iota\geq 3$. Also, let $m$ and $h$ satisfy $n/m^4=o(1)$, $n h^p/m^3\rightarrow \infty$, and $n h^{2\rho}=o(1)$. Then, (\ref{eq: IFR for betahat}) holds true. \\ 

\noindent If moreover Assumption \ref{ass: identification}\textcolor{blue}{(b)} holds, then (\ref{eq: rate for phihat in the PL model}) holds true.
\end{enumerate}

\end{thm}
Part (i) of Theorem \ref{prop: asy norm of betahat and convergence rates of phihat in the PL model} obtains the asymptotic normality of $\sqrt{n}(\widehat \beta - \beta_0)$ under Assumption \ref{ass: source condition on A x}. Part (ii) shows that it is possible to avoid Assumption \ref{ass: source condition on A x} when the degree of smoothness of $\phi_0$ is sufficiently large in terms of the source condition ($\iota\geq 3$ in Assumption \ref{ass: source condition}). This possibility arises thanks to the Landweber-Fridman scheme which can exploit orders of smoothness of $\phi_0$ larger than 2. This would not have been possible if we used a Tikhonov regularization scheme, since it cannot exploit order of smoothness of $\phi_0$ larger than 2.\footnote{Theorem \ref{prop: asy norm of betahat and convergence rates of phihat in the PL model}(i) imposes the same conditions as Theorem \ref{prop: asynorm of the innder product} on $m$, $h$, and $\rho$. We have discussed these conditions  below Theorem \ref{prop: asynorm of the innder product}. To avoid Assumption \ref{ass: source condition on A x}, Theorem \ref{prop: asy norm of betahat and convergence rates of phihat in the PL model}(ii) also requires different conditions on $m$ and $h$ with respect to Theorem \ref{prop: asy norm of betahat and convergence rates of phihat in the PL model}(i). If $h\sim n^{-\gamma}$ and $m\sim n^\alpha$, then the conditions of Theorem \ref{prop: asy norm of betahat and convergence rates of phihat in the PL model}(ii) will be satisfied for $1/4<\alpha<1/3$, $\gamma<(1-3\alpha)/p$, and $\rho>1/(2\gamma)$. Finally, notice that the conditions on $m$, $h$, and $\rho$ in Theorem \ref{prop: asy norm of betahat and convergence rates of phihat in the PL model}(i) or Theorem \ref{prop: asy norm of betahat and convergence rates of phihat in the PL model}(ii) imply the conditions in Corollary \ref{corollary: consistency of phihat}. }  \\
By the above theorem, the asymptotic normality of $\sqrt{n}(\widehat \beta - \beta_0)$ does not necessarily require the injectivity of $A_Z$ and hence the completeness of the distribution of $Z$ conditional on $W$. Thus, the asymptotic normality of $\sqrt{n}(\widehat \beta - \beta_0)$ holds regardless of whether $\phi_0$ is identified. \cite{chen2021robust} also obtains the asymptotic normality of the slope coefficients of a partly linear IV model without necessarily relying on the completeness assumption. While \cite{chen2021robust} uses a series  estimator based on first-stage regressions on the IVs, we are here using a Landweber-Fridman regularization without requiring first-stage regressions or first-stage smoothing on the IVs.

Notice that when $A_Z$ is injective $U^\perp=U$ almost surely, so that the error $U$ will appear in the asymptotic variance of $\sqrt{n}(\widehat \beta - \beta_0)$.\footnote{From the Direct Sum Theorem, $\pi$-almost everywhere we have $\phi_0=\phi_0^\perp+P_{\mathcal{N}(A_Z)}\phi_0$, where $P_{\mathcal{N}(A_Z)}\phi_0$ represents the projection of $\phi_0$ onto $\mathcal{N}(A_Z)$.   When $A_Z$ is injective, $\mathcal{N}(A_Z)=\{0\}$, so that $P_{\mathcal{N}(A_Z)}\phi_0=0$ and $\phi_0=\phi_0^\perp$ $\pi$-almost everywhere. This implies that $\int |\phi_0(z)-\phi_0^\perp(z)|f_Z(z)dz=\int |\phi_0(z)-\phi_0^\perp(z)|\cdot |f_Z(z)/\pi(z)|\,\pi(dz)$ $\leq \|\phi_0-\phi_0^\perp\| [\int |f_Z(z)/\pi(z)|^2 \pi(dz)]^{1/2}=0$. Thus, $\phi_0(Z)=\phi_0^\perp(Z)$ almost surely and hence $U^\perp=U$ almost surely. }  
Thanks to the root-$n$ convergence rate of $\widehat{\beta}$, the convergence rate of $\widehat{\phi}$ is not affected by the preliminary estimation of $\beta_0$. \\
From a practical standpoint, although $\sqrt{n}(\widehat{\beta}-\beta_0)$ is asymptotically normal, the covariance matrix of the asymptotic distribution has an intricate expression. So, in practice to test hypotheses on $\beta_0$ we suggest bootstrapping the statistic $\sqrt{n}(\widehat{\beta}-\beta_0)$ according to the pairwise scheme. Although we do not provide a formal proof for the validity of the pairwise bootstrap, its employment can be informally justified by the asymptotic normality of $\sqrt{n}(\widehat \beta-\beta_0)$.
In our simulation study contained in the Supplementary Material, we show that bootstrapping the statistic  $\sqrt{n}(\widehat \beta-\beta_0)$ yields a good behavior for the Wald test in finite samples.

\section{Concluding Remarks}
We have studied an estimation method for partly linear IV models that does not smooth on the IVs and is based on the Landweber-Fridman regularization. We have obtained the convergence rate of the nonparametric estimator and the asymptotic normality of the parametric estimator. This asymptotic normality result does not rely on the completeness assumption.  \\
An area for further investigation is the connection between the proposed approach based on the continuum of moments and the recent development on locally robust estimators, see \cite{chernozhukov2022locally}. It would also be interesting to study the efficiency of the proposed approach for semiparametric IV models. Finally, the pointwise  asymptotic normality of the nonparametric estimator is a further topic for future research.

\bigskip\noindent


\noindent\textbf{Supplementary information}
The Supplementary Material contains the details about the implementation of our estimators, the Monte Carlo simulations, and the proofs of the theorems.

\bigskip\noindent

\noindent\textbf{Acknowledgments}  We thank two anonymous referees for their comments that helped to improve the paper. We thank Pascal Lavergne for inspiring discussions. We are also grateful to Juan Carlos Escanciano, Ingrid Van Keilegom, Jad Beyhum, Valentin Patilea, and Xavier d'Haultfoeuille for their helpful comments. Jean-Pierre Florens acknowledges funding from the French National Research Agency (ANR) under the Investments for the Future program (Investissements d'Avenir, grant ANR-17-EURE-0010). 



\bigskip
\bibliographystyle{ecta}
\footnotesize
\renewcommand{\baselinestretch}{1}
\bibliography{sn-biblio}
\normalsize


\end{document}